# Diminished Short Channel Effects in Nanoscale Double-Gate Silicon-On-Insulator Metal-Oxide-Semiconductor Field-Effect-Transistors due to Induced Back-Gate Step Potential

M. Jagadesh Kumar[*] and G. Venkateshwar Reddy

Department of Electrical Engineering,
Indian Institute of Technology,
Hauz Khas, New Delhi 110016, India

## Abstract

In this letter we discuss how the short channel behavior in sub 100 nm channel range can be improved by inducing a step surface potential profile at the back gate of an asymmetrical double gate (DG) Silicon-On-Insulator (SOI) Metal-Oxide-Semiconductor Field-Effect-Transistor (MOSFET) in which the front gate consists of two materials with different work functions.

[*] E-mail address: mamidala@ieee.org

# 1. Introduction

Channel length reduction to sub-100 nm lengths leads to two important short channel effects: (i) threshold voltage roll-off and the drain induced barrier lowering (DIBL). A number of solutions have been proposed to minimize the short channel effects (SCEs) such as thin-body single material gate (SMG) silicon-on-insulator (SOI) metal-oxide-semiconductor field-effect-transistors (MOSFETs), double material gate (DMG) SOI MOSFETs and double gate (DG) SOI MOSFETs.[1,2]. Recently, it has been shown that by replacing the single material gate of the SOI MOSFETs by a double material gate, the short-channel effects can be further controlled [3]. In the DMG SOI MOSFET, the gate is made of two materials with different workfunctions to introduce a potential step in the channel region [4]. As a result, the channel potential minima on the source side is unaffected by the drain voltage variations resulting in improved SCEs in the DMG SOI MOSFETs [5,6].

The primary objective of this letter is to demonstrate that if the double material gate concept is applied to the front gate of the DG SOI MOSFET, due to a strong coupling between the front and back gates, a step potential can be induced at the back gate even without having to use a double material for the back gate. Consequently, due to the induced step potential at the back gate, the double material double gate (DMDG) SOI MOSFET will have significantly improved gate control over the channel region leading to the diminished SCEs. Using two-dimensional simulation [7], we demonstrate that even for sub-100 nm channel lengths, the DMDG SOI MOSFET exhibits the desirable threshold voltage roll-up and diminished DIBL as compared with the DG SOI MOSFET due to the induced step potential at the back gate.

# 2. Simulation Results

Schematic cross-sectional views of both the asymmetrical DG and DMDG SOI MOSFETs implemented in the two-dimensional device simulator MEDICI [8] are shown in

Figure 1. The asymmetrical DG SOI MOSFET consists of a front $p^+$ poly gate and a back $n^+$ poly gate. In the case of the DMDG SOI MOSFET, the front gate consists of dual materials M1 ($p^+$ poly) and M2 ($n^+$ poly) of lengths $L_1$ and $L_2$ respectively, while the back gate is an $n^+$ poly gate. In our simulations, the silicon thin-film thickness ($t_{si}$) is 12 nm, the front-gate oxide ($t_f$) and the back-gate oxide ($t_b$) thickness is 2 nm. The doping in the p-type body and n+ source/drain regions is kept at $1 \times 10^{15}$ and $5 \times 10^{19} cm^{-3}$ respectively.

Figure 2 shows the surface potentials at the front and the back gates of the DG and DMDG SOI MOSFETs along the channel. It can be observed that the DMDG structure exhibits a step in the surface potential profile at the front gate as well as at the back gate. The step, which is quite substantial in the front gate surface potential, occurs because of the difference between the work functions of $n^+$ poly and $p^+$ poly [3]. Though the back gate in DMDG structure consists of a single material ($n^+$ poly), an induced potential step can be seen at the back gate when the silicon film thickness is thin due to the coupling between the front channel and back channel. The back gate surface potential profile plays a dominant role in deciding the threshold voltage of an asymmetrical DG SOI MOSFET and due to the presence of the induced step profile at the back gate, the short channel behavior of this structure is expected to improve as discussed below.

The short channel behavior of the DG and the DMDG SOI MOSFETs is evaluated based on the threshold voltage and the DIBL variation with channel lengths. Figure 3 shows the threshold voltage of the DG and the DMDG structures for different channel lengths below 100 nm with the length of the $p^+$ poly ($L_1$) in the DMDG SOI MOSFET fixed at 50 nm. It can be clearly observed that the threshold voltage decreases with smaller channel lengths for the DG structure while a threshold voltage roll-up can be observed for the DMDG structure. The other important parameter which characterizes the short channel performance is the drain induced barrier lowering (DIBL). Figure 4 shows the DIBL for the DG and the DMDG SOI

MOSFETs for different channel lengths. It can be observed that the DIBL, which is the difference between the linear threshold voltage ($V_{th,lin}$) and the saturation threshold voltage ($V_{th,sat}$) is far less for DMDG structure when compared with that of the DG structure. The reduced DIBL in the DMDG SOI MOSFET is because of the screening of the drain potential by the induced step surface potential profile at the back gate. The step profile ensures that the source side of the channel is screened from the drain potential variations so that the surface potential minima at the source end remains effectively unchanged resulting in a reduction in DIBL.

## 3. Conclusions

We have shown that the coupling between the front gate and the back gate in a modified thin film double gate SOI MOSFET induces a step in the back gate surface potential profile. This helps in attaining the desirable threshold voltage roll-up and reduced DIBL with decreasing channel lengths when compared with the DG SOI MOSFET.

**Figure Captions**

Figure 1   Cross-sectional view of (a) DG-SOI MOSFET and (b) DMDG-SOI MOSFET.

Figure 2.   Surface potential profiles at the front gate and the back gate for the DG and the DMDG-SOI MOSFETs with a channel length $L = 0.1$ μm ($L_1 = L_2 = 0.05$ μm).

Figure 3.   Threshold voltage of the DG and the DMDG SOI MOSFETs for different channel lengths ($L_1$ fixed at 0.05 μm).

Figure 4.   DIBL of the DG and the DMDG SOI MOSFETs for different channel lengths with $L_1 = L_2$.

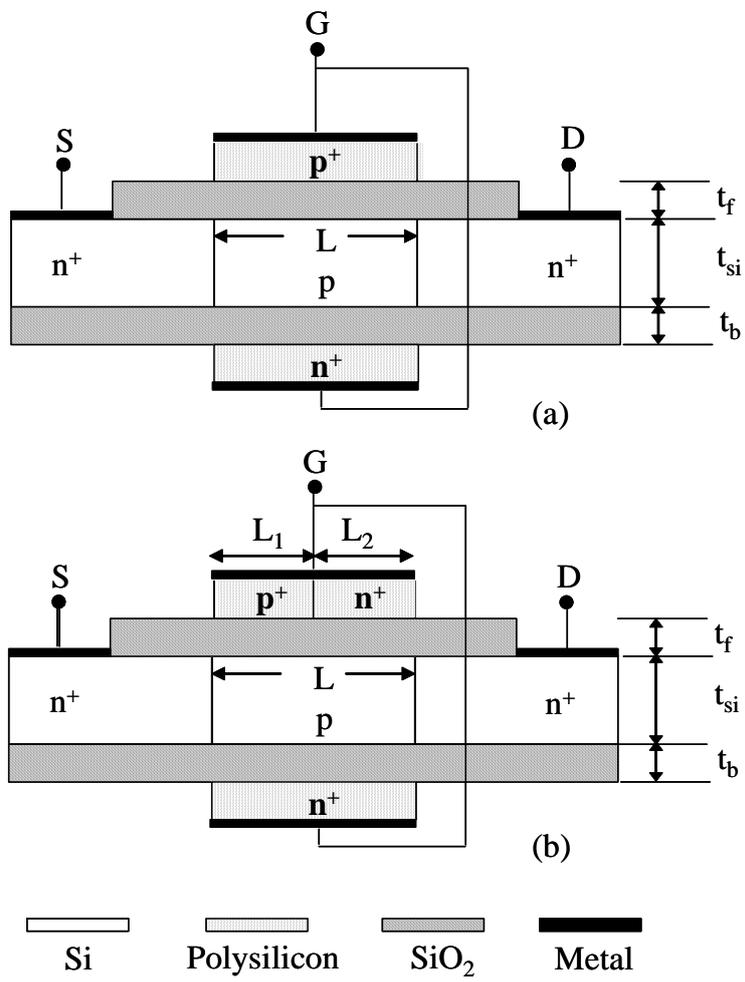

Fig. 1

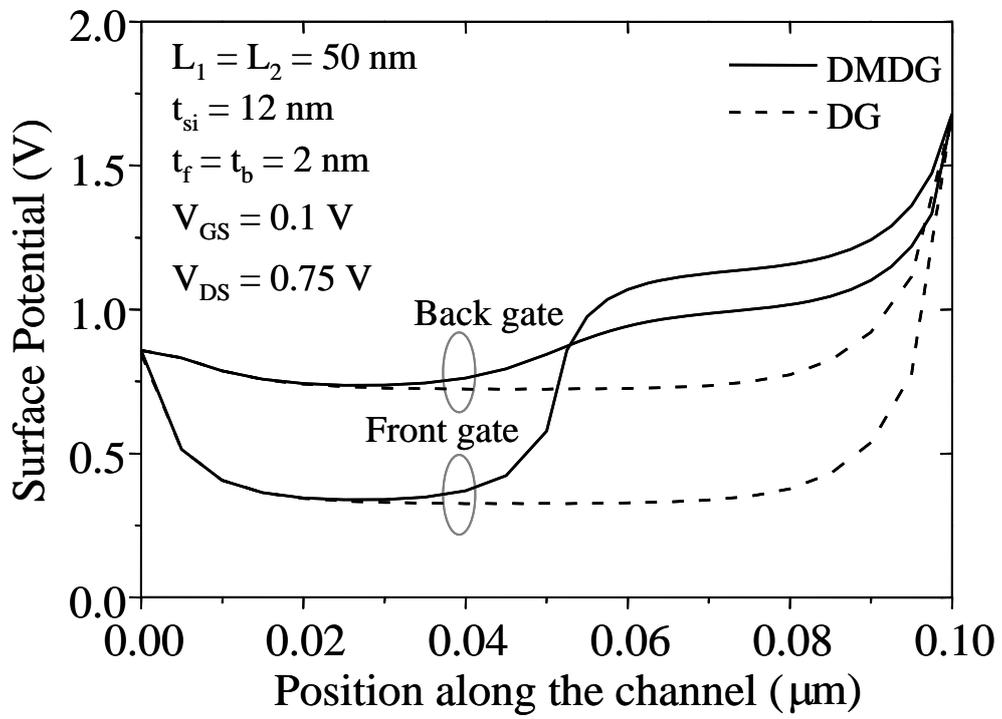

Fig. 2

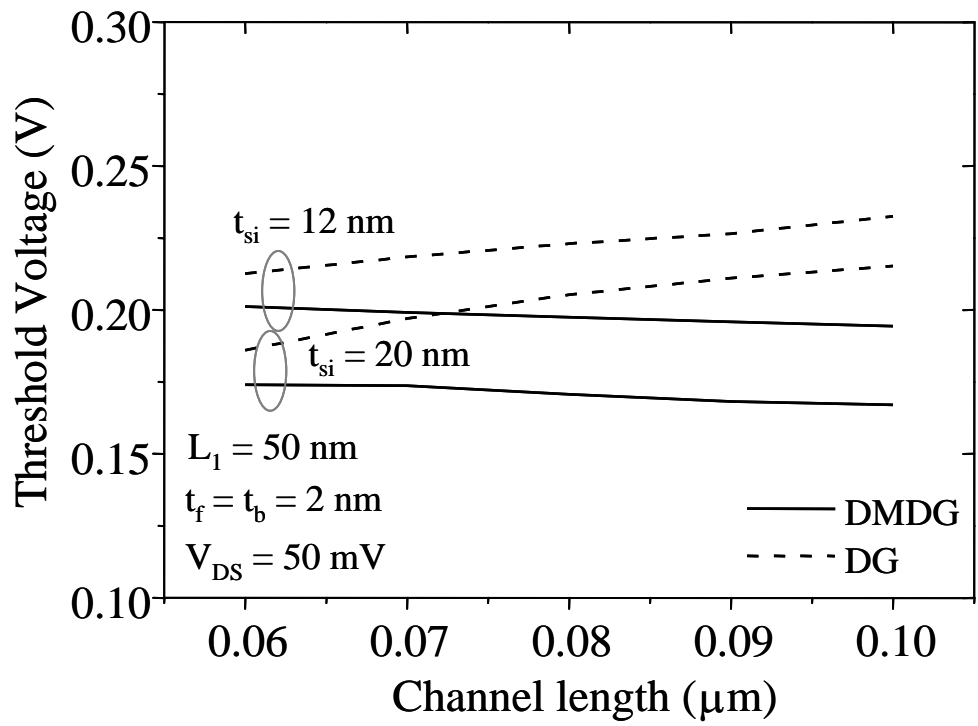

Fig. 3

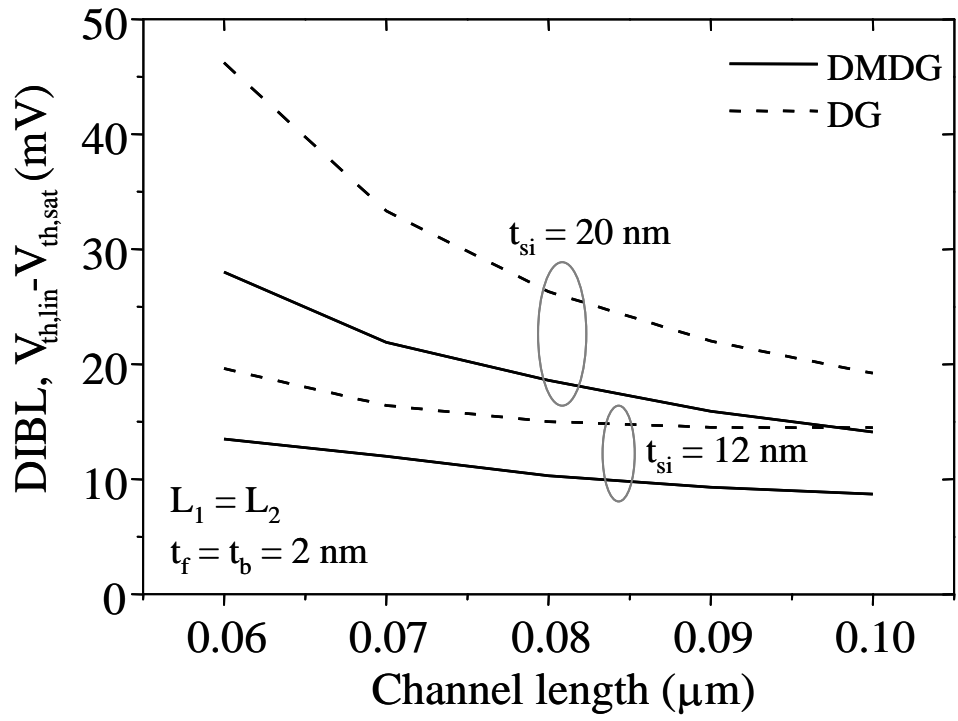

Fig. 4